# 'It Has to Go Down A Little, In Order to Go Around'- Following Feynman on the Gyroscope

Svilen Kostov and Daniel Hammer
*Georgia Southwestern State University*

In this paper we show that with the help of accessible, teaching quality equipment, some interesting details of the motion of a gyroscope, typically overlooked in introductory courses, can be measured and compared to theory. We begin by deriving a simple relation between the *asymptotic dip angle* of a gyroscope released from rest and its *precession velocity*. We then describe an experiment which measures these parameters. The data gives excellent agreement with the model. The idea for this project was suggested by the discussion of gyroscopic motion in The Feynman Lectures on Physics. Feynman's conclusion (stated in colloquial terms and quoted in the title) is confirmed and, in addition, conservation of angular momentum, which underlies this effect, is quantitatively demonstrated.

**Introduction**

In the history of physics teaching, The Feynman Lectures on Physics[1] (FLP) are perhaps better known as an expression of the brilliant and idiosyncratic mind of their famous author, than the modern pedagogical tool that they were intended to be at the time of their publication. The legendary status which these books have enjoyed since then is due, in part, to the fact that they have been a source inspiration to the many instructors and teachers (and even students) who have used them to discover Feynman's imaginative and unconventional perspective on some rather familiar topics in elementary physics.

The treatment of the spinning top or gyroscope is one such example. It is interesting, in part, for what has become known about Feynman's early fascination with this phenomenon, stemming from an observation of a wobbling plate in the cafeteria[2]. The discussion in the FLP, very importantly, addresses the apparent paradox of how is precessional motion at all possible given the fact that the external torque is always in the horizontal plane? In the end, Feynman dispels the "miracle" of the purely precessional motion by bringing *nutation* (the wobbling part) into the picture as the mechanism by which some angular momentum is transferred from the horizontal to the vertical direction. He concludes that this forces the spin axis to undergo a small dip. The phenomenon itself is familiar to anyone who has taken the time to play with a simple gyroscope, bicycle wheel or gimbal type. What may not be so obvious is that this follows from conservation of total angular momentum, and that it can be quantified and measured in a fairly simple and straightforward manner.

**A Simple Model for the Dip of a Spinning Top Released From Rest**

We approximate the gyroscope by a heavy thin disk on a massles axle (Fig. 1a). It is prepared with a large angular velocity (at least several hundred rpm, so that it is a true gyroscope) along the horizontal axis *x* and then released from rest. It begins to undergo both precession and nutation. The nutation decays due to friction in the bearings and eventually (asymptotically) the gyroscope settles to a constant precession with its axis having dipped slightly below the initial level at an angle of $\Delta\theta$. (Fig. 1b). It is clear that for small dip angles, the original spin angular momentum vector retains its magnitude; $L \approx L_0$ (Fig. 1c), where $L_0$ is the initial and $L$ is the final angular momentum along the spin axis.

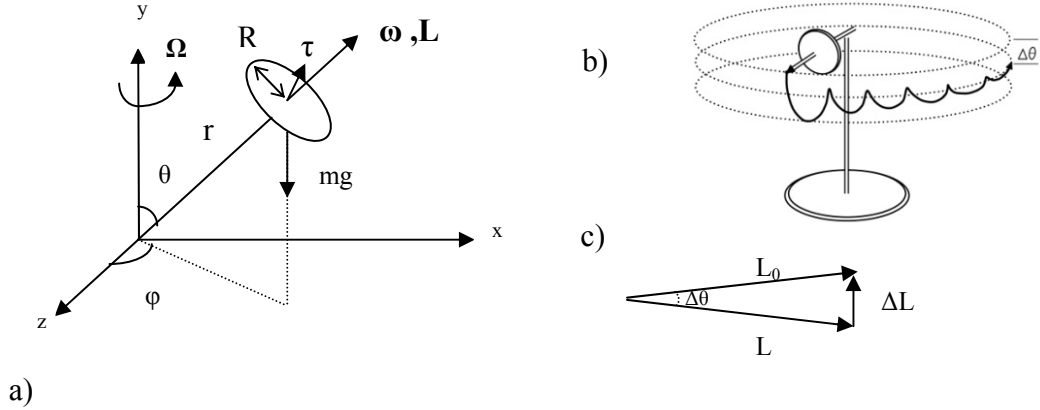

a)

**Figure 1. a) Basic parameters in the torque induced precession of gyroscope. The external torque τ is in the *xz* plane.
b) Precession and decaying nutation. The dip angle, Δθ, is indicated.
c) Relation between initial and final spin angular momentum, after nutation has died out.**

A small vertical component, **ΔL,** arises as the difference between the two. For the magnitudes we have:
$$\Delta L \approx L_0 \Delta\theta \tag{1}$$

The corresponding angular velocity for this motion (about the vertical axis *y*) is identified with the precessional velocity, derived in introductory texts by examining the torque acting on the falling disk and the induced incremental change in the angular momentum vector in the horizontal plane. It is inversely related to the spin angular velocity and given by the well known formula

$$\Omega = \frac{mgr}{\lambda_3 \omega}, \tag{2}$$

where $m$ is the mass of the disk, $r$, the distance from the vertical axis, $\omega$ the spin angular velocity about the horizontal axis, and $\lambda_3$ the moment of inertial of the disk about the horizontal spin axis. For a thin disk,

$$\lambda_3 = \frac{1}{2}mR^2, \tag{3}$$

where R is the radius of the disk. It is important to note that in the conventional derivation of Eq. (2) one assumes a continuous change of the angular momentum (direction) in the horizontal but no change in the vertical plane. It should be clear, however, that the precession velocity in this equation is precisely that of **ΔL**, of Eq. (1), hence, vertical angular momentum somehow comes about! How? The answer has to be that it is "borrowed" from the initial vector during the nutation; hence $L_0$ dips and becomes (negligibly) shorter. The dip is directly measurable. We should point out that in this paper we are not addressing the mechanism of nutation. Feynman and others[4] have attempted to explain it in elementary terms. We focus on the asymptotic, end result.

Rewriting angular momentum in terms of angular velocity and moment of inertia, Eq. (1) becomes:

$$\Omega \lambda_1 = \omega \lambda_3 \Delta \theta, \qquad (4)$$

where $\lambda_1$ is the moment of inertia about the vertical axis. For a thin disk it can be computed using the parallel axis theorem to be

$$\lambda_1 = \frac{1}{4} mR^2 + mr^2 \qquad (5)$$

If in Eq. (2) we express $\omega$ in terms of $\Omega$ and substitute it into Eq. (4), we obtain a simple relation between the dip angle (the "going down") and the precession velocity (characterizing the "going around"):

$$\Delta \theta = \frac{\lambda_1 \Omega^2}{mgr}, \qquad (6)$$

hence, in the case of large spin velocity, the dip angle should be proportional to the square of the precession velocity if total angular momentum is to be conserved.

**An Experiment to Test the Model**

For the purpose of this experiment, we have used a demonstration gyroscope from PASCO (Fig. 2). We added two PASCO digital angle sensors which allow the simultaneous measurement of $\varphi(t)$ and of $\theta(t)$. This allows for a direct determination of $\omega$ as the $\varphi(t)$ sweep rate. The gyroscope disk was accelerated to exact initial spin velocities using a battery powered drill with a rubber tip. The values were calibrated by observing the motion of a small dot on the disk while it was illuminated by a digital strobe light set at specific frequencies. Aliasing effects were taken into account in determining when the exact rotational frequency was reached, so that the top could be

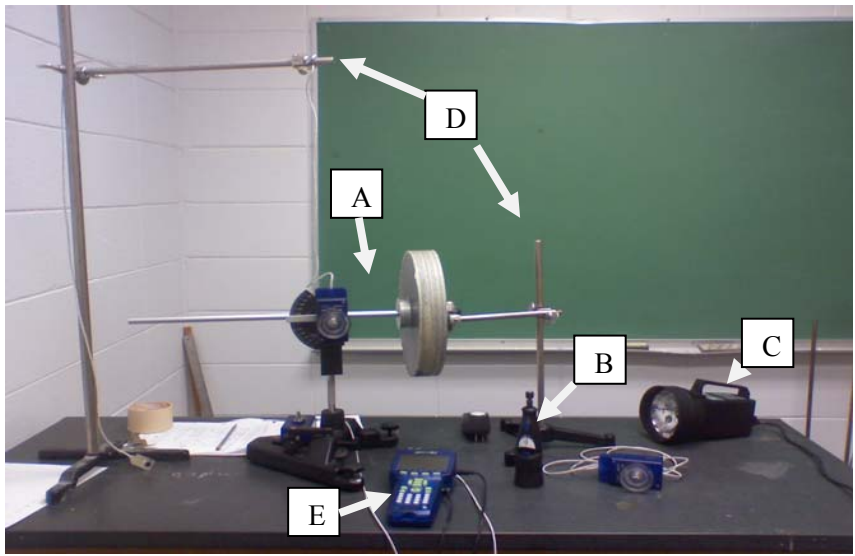

Figure 2. The experimental setup:

A. Demonstration gyroscope (PASCO) with two digital rotational sensors
B. Mini drill (DREMEL) with rubber head to accelerate disk to target rpm values
C. Digital stroboscope (EXTEC Instruments) to measure initial spin rate by observing aliasing pattern of spot on disk
D. Supporting stands
E. Data acquisition device ('Explorer', PASCO)

released. The data, at a sample rate of 1/50 s was collected with a hand held Explorer device from PASCO.

Data for the precession angle $\varphi(t)$ and corresponding nutation angle $\theta(t)$ was obtained for seven different spin velocities: 400 rpm, 600 rpm, 800 rpm, 1000 rpm, 1200 rpm, 1400rpm, and 1600 rpm. In all the runs, the initial precessional velocity was $\dot{\varphi}(0) = 0$ and the initial azimuthal angle, $\theta(0) = \frac{\pi}{2}$. Plots of some of the experimentally obtained curves, $\varphi(\theta)$, are given in Fig. 3. It is evident that at high spin values, the attenuation of the nutation is rapid and the top quickly settles to its dip angle. At lower spin velocities, the nutation persists over many revolutions of the disk, so we have taken $\Delta\theta$ to be from the horizontal to the midpoint of each full dip (averaged).

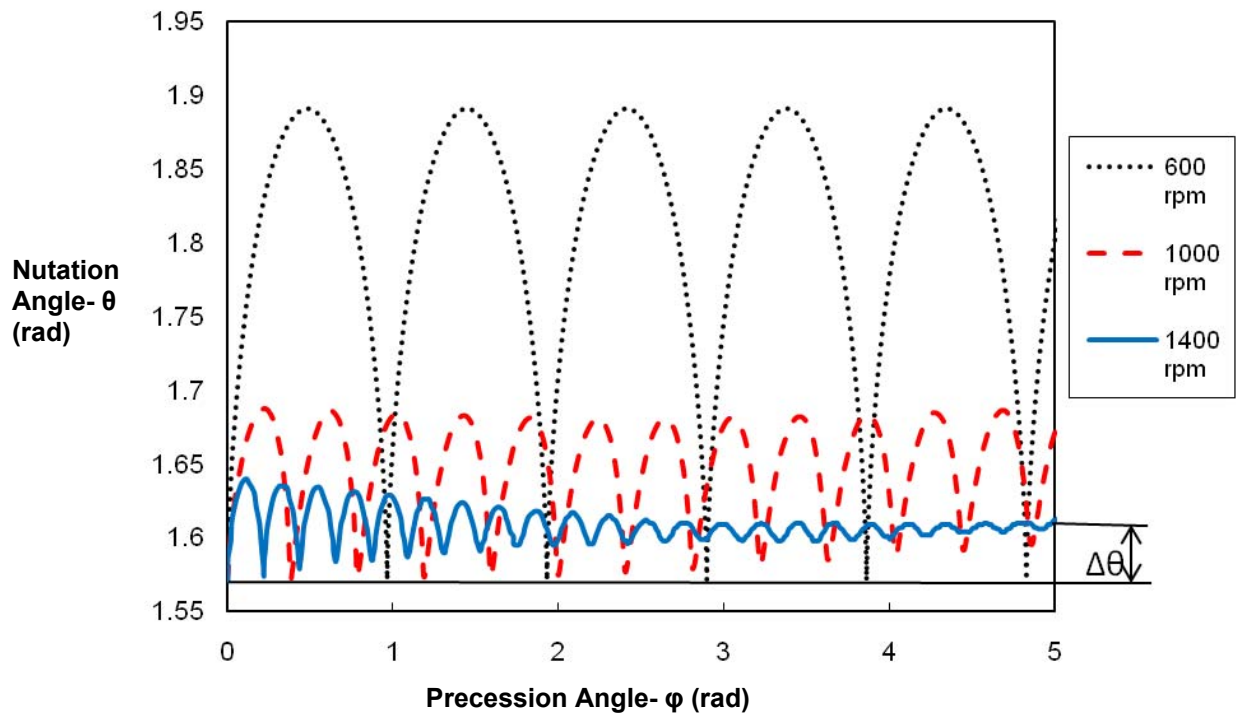

**Figure 3. Experimental φ vs. θ curves for three different spin velocities. Attenuation is significant at the highest spin value. Dip angle for this (1400 rpm) case is shown.**

It is important to note that the dip of the spin axis is not the result of energy (and angular momentum) loss at the bearings during the attenuation of the nutation. It can easily be shown that such energy loss, for example, over a full precession period, is insignificant compared to the initial kinetic energy of the top, and can therefore be neglected.

A plot of the dip angle $\Delta\theta$ against $\Omega^2$, corresponding to the different spin velocities (in decreasing order from left to right), shows a linear dependence, as expected (Fig. 4). The slope of the graph agrees with the values of the parameters in Eq. (6) within two percent, which is consistent with the error propagation.

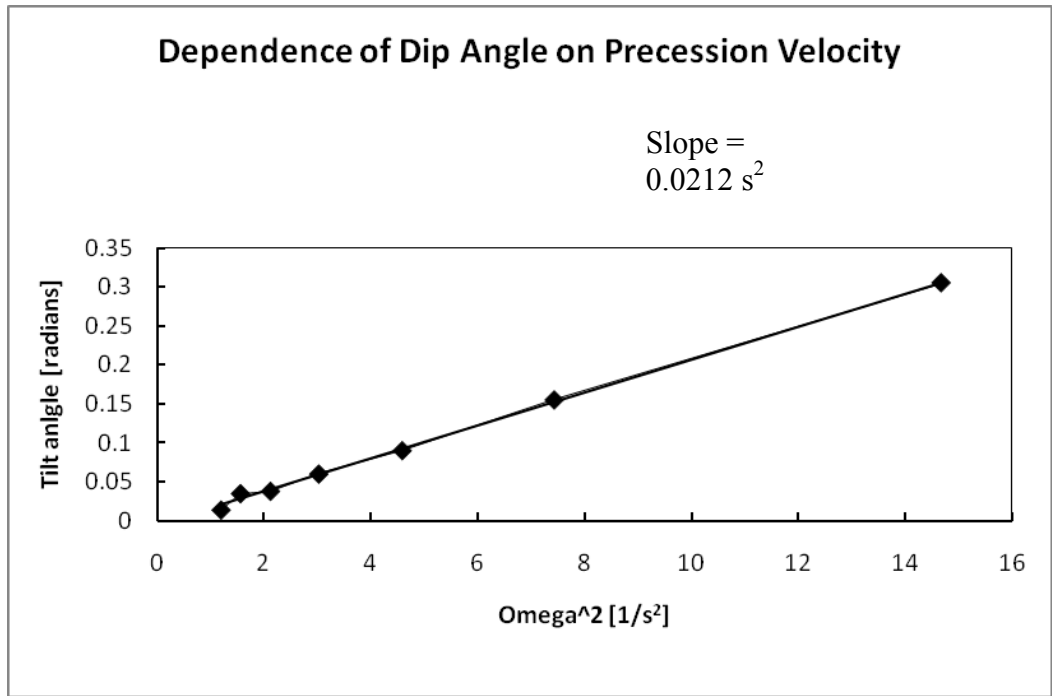

**Figure 4. Dip angle ($\Delta\theta$) plotted against the square of the precessional velocity ($\Omega^2$). The slope is in excellent agreement with $\dfrac{\lambda_1}{mgr}$ where $\lambda_1$ = 0.052±0.001 kg m$^2$, m = 1.700 ±0.010 kg, r = 0.145 ±0.002 m, and g = 9.81 m/s$^2$**

**Conclusion**

The work described in this paper was suggested to us from a rereading of the Feynman Lectures on Physics. We consider it to be an interesting and seamless extension of the conventional analysis of gyroscope motion at the introductory physics level. It can also serve as a bridge between the elementary and more exact, detailed treatments of this phenomenon, suitable for higher level mechanics courses[5, 6, 7]. It is also one of the few lab projects at the undergraduate level, which we are aware of, for exploring angular momentum.

**Acknowledgements**

The authors would like to thank the Department of Geology and Physics at Georgia Southwestern State University for supporting this project. We also wish to thank